\documentstyle[aps,prl,epsf,floats]{revtex} 
\def\lsim{\mathrel{\vcenter{\hbox{$<$}\nointerlineskip\hbox{$\sim$}}}}

\draft

\begin{document}
\twocolumn[\hsize\textwidth\columnwidth\hsize\csname
@twocolumnfalse\endcsname 

\preprint{UCLA/01/TEP/11\\ VAND-TH-01-06}

\title{
Neutrino cross sections and future observations of
ultrahigh-energy cosmic rays
}

\author{Alexander Kusenko$^{1,2}$ and Thomas Weiler$^3$}
\address{$^1$Department of Physics and Astronomy, UCLA, Los Angeles, CA
90095-1547 \\ $^2$RIKEN BNL Research Center, Brookhaven National
Laboratory, Upton, NY 11973 \\
$^3$ Department of Physics and Astronomy, Vanderbilt University,
Nashville, TN 37235
}


\maketitle
             
\begin{abstract}

We show that future detectors of ultrahigh-energy cosmic-ray neutrinos will
be able to measure neutrino-nucleon cross section, $\sigma_{\nu N}$, at
energies as high as $10^{11}$GeV or higher.  We find that the flux of up-going
charged leptons per unit surface area 
produced by neutrino interactions below the surface is {\sl
inversely} proportional to $\sigma_{\nu N}$.  
This contrasts with the rate of horizontal air
showers (HAS) due to neutrino interactions in the atmosphere which is
proportional to $\sigma_{\nu N}$.
Thus, by comparing the HAS
and up-going air shower (UAS) rates, the neutrino-nucleon cross section can
be inferred.   Taken together, up-going and horizontal rates ensure a
healthy total event rate,  
regardless of the value of $\sigma_{\nu N}$.

\end{abstract}

\pacs{PACS numbers: 13.85.Tp; 95.55.Vj; 95.85.Ry; 96.40.Pq \hfill 
UCLA/01/TEP/11; VAND-TH-01-06}

]


\renewcommand{\thefootnote}{\arabic{footnote}}
\setcounter{footnote}{0}

%
Detection of ultrahigh-energy (UHE) neutrinos is one of the important
challenges of the next generation of cosmic ray detectors.  Their discovery
will mark the advent of UHE neutrino astronomy, allowing the mapping on the
sky of the most energetic, and most distant, sources in the Universe.  In
addition, detection of UHE neutrinos may help resolve the
puzzle~\cite{puzzle} of cosmic rays with energies beyond the
Greisen-Zatsepin-Kuzmin cutoff~\cite{gzk} by validating 
Z-bursts~\cite{Zburst}, topological defects~\cite{TDs}, superheavy
relic particles~\cite{SMPs}, new strong-interactions~\cite{nuSI}, {\it etc.}

Several approved and proposed experiments plan to detect
UHE neutrinos by observation of the nearly horizontal air
showers (HAS) in the Earth's atmosphere resulting from $\nu$-air
interactions.  
The expected rates are proportional to the neutrino-nucleon cross
section.  Calculations of this cross section at $10^{20}$~eV necessarily
use an extrapolation of parton distribution functions and Standard Model
(SM) parameters far beyond the reach of experimental data. 
The resulting cross section at $10^{20}$~eV is $\sim 10^{-31}{\rm cm}^2$
\cite{UHEnusig}.  It has recently been argued that the
extrapolated neutrino cross section may be too high~\cite{Dicus} 
at energies above about $3\times 10^{17}$eV.  If the cross section 
is lower, then the event rate for
neutrino-induced HAS is reduced by the same factor.  This reduction would 
compromise the main detection signal that has been proposed for
UHE neutrino experiments.
On the other hand, the extrapolated cross-section may be too low,
for it ignores possible contributions from new physics.

We will show, however, that regardless of possible theoretical
uncertainties in the cross section, the future experiments can observe the
UHE neutrinos.  In fact, a smaller cross section
would offer a double advantage for the planned experiments.  First, with a
new search strategy (described below), the neutrino event rate with a small
cross section is actually
larger than the HAS rate with a large cross section.  Second, the future
detectors can also measure the neutrino cross section at energies far
beyond those achievable in collider experiments.  The first advantage is a
boon for neutrino astronomy, while the second provides important
information for particle physics.  
Here we will take the value of the cross
section to be a free parameter with a wide range of values.

This study is motivated in part by a recent analysis of upward events
by Feng et al. \cite{Feng}.  The emphases in the two papers are quite
different.

In addition to HAS, proposed cosmic ray experiments can also observe
up-going air showers (UAS) initiated by muon and tau leptons 
produced by neutrinos interacting just below the surface of the Earth,
and may possibly observe the fluorescence signal from 
up-going charged muon and tau leptons (UCL) themselves.
Prior estimates for the rate of ``earth-skimming'' events have used the
extrapolated neutrino cross section~\cite{Feng,Domokos}.  
A smaller value of
this cross section reduces the shadowing of UHE neutrinos by the Earth.
Therefore, the neutrino angles with respect to horizon need not be so
``skimming.'' 
More importantly, the expected rate of UCL and UAS 
may (i) be larger, and (ii) depend on the cross section.

Indeed, a lower cross section {\em increases} the UCL rate 
per surface area 
as $\sigma_{\nu N}^{-1}$ as long as the neutrino absorption 
mean free path (MFP) in Earth
is small in comparison with the Earth's radius, $ R_\oplus$;  
i.e., for $\sigma_{\nu N}  \stackrel{>}{_{\scriptstyle \sim}} 
 2\times 10^{-33}{\rm cm}^2$
the UCL event rate is 
proportional to $F_\nu/\sigma_{\nu N}$, as shown in Fig.~1.  
This inverse dependence of the UCL rate on the cross section is to be
contrasted with the rate of HAS events resulting from $\nu$-air
interactions, which decreases as $\sigma_{\nu N}$ decreases.  
Furthermore, we find that
for small but possible values of the cross section, the UAS rate can 
exceed the HAS rate by several orders of magnitude,
as displayed in Fig. 2.  

There is a firm 
prediction for a flux $F_\nu $ of GZK neutrinos
in the energy range $10^{15}$ to $10^{20}$~eV, based on
the observed flux of UHECR protons at the GZK limit. 
This flux is expected to peak in the decade 
$10^{17}$ to $10^{18}$~eV for uniformly-distributed 
proton sources, and around $10^{19}$~eV for ``local'' sources 
within $\sim 50$~Mpc of earth~\cite{ESS01}.
If the GZK flux is the dominant source  of UHE neutrinos, 
then one can use the predicted GZK flux value and 
either the inverse relation between the UCL rate and $\sigma_{\nu N}$
(shown in Fig.~1), or the linear relation between the HAS rate and 
$\sigma_{\nu N}$ to infer from future data the value of
the neutrino cross section at ultrahigh energies.
\begin{figure}[t]
\centering
\leavevmode 
\epsfxsize=8cm 
\epsfbox{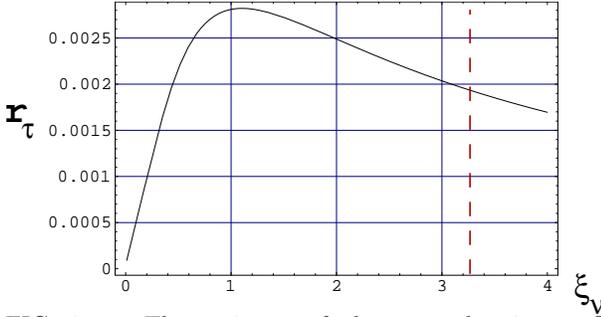}
\caption[fig. 1]{\label{fig1} 
The ratio {$r_\tau $} of the upward going $\tau$ flux to the incident tau
neutrino flux $F_{\nu_\tau} $ as a function of {$\xi_\nu=\lambda_\nu
/R_\oplus =1/(\sigma_{\nu _N} n R_\oplus) $}, from eq.\ (5), with fixed
$\lambda_\tau/R_\oplus =3.5\times 10^{-3}$, appropriate for events initiated by
$\sim 10^{20}$~eV neutrinos.  Here $n$ is the mean nucleon number density.
Assuming a monotonic cross section dependence on $\sqrt{s}$, the value of
$\xi $ is limited from above by the HERA measurements, as shown by a
vertical dashed line. }
\end{figure} 
In addition, a comparison of the UAS and HAS rates,
or a measurement of the angular distribution of UCL/UAS 
events may allow an experiment to determine $\sigma_{\nu N}$
independent of $F_\nu$, as discussed below.
More promising for $\sigma_{\nu N}$ determination are
possible neutrino fluxes at and above $10^{20}$~eV 
given, {\em e.g.}, in \cite{Zburst,Zburst1}.

Let us now estimate the rate of upward showers. 
UHE neutrinos are expected to arise from pion and subsequent muon decay.
The initial flavor composition is therefore
$\nu_\mu$ and $\nu_e$ with a ratio 2:1.
These flavors
oscillate and eventually decohere during their Hubble-time journey.
The resulting neutrino state includes a 
$\nu_\tau$ fraction $\frac{2}{3}\sum_j |U_{\mu
j}|^2\,|U_{\tau j}|^2 + \frac{1}{3}\sum_j |U_{e j}|^2\,|U_{\tau j}|^2$,
where $U_{\alpha j}$ are the mixing elements relating the neutrino mass and
the flavor bases.  
If $|U_{\tau 3}|\simeq |U_{\mu 3}|$ are large, as inferred from 
the Super-Kamiokande data, then 
the oscillations nearly equalize the
number of UHE neutrinos of each flavor~\cite{decohere}.
In particular, $F_{\nu_\tau}\approx\frac{1}{3}F_\nu$ is expected.
The energy-loss MFPs, $\lambda_{\tau}$ and $\lambda_\mu$, for
taus and muons to lose 
a decade in energy are 11~km and 1.5~km, respectively, in surface rock
with density $\rho_{sr} = 2.65\,{\rm g/cm}^3$,
and 2.65 times longer for lepton trajectories passing
through ocean water.  
Tau and muon decay MFPs are long above 
$10^{18}$~eV: 
$c\tau_\tau=490\,(E_\tau / 10^{19}{\rm eV})$~km,
and $c\tau_\mu \sim 10^8\,c\tau_\tau$ 
for the same lepton energy.
Because the energy-loss MFP for a $\tau$ produced in rock or
water is much longer than that of a muon, the produced taus have a much
higher probability to emerge from the Earth and to produce an atmospheric
shower.  Thus, the
dominant primary for initiation of UAS events is the tau neutrino.  
In what follows we focus on tau neutrinos incident at $10^{20}$~eV.

Let us consider an incident tau neutrino whose trajectory 
cuts a chord of length $l$ in the Earth.  
The probability for this neutrino to reach a
distance $x$ is $P_\nu (x) =e^{-x/\lambda_\nu}$, where
$\lambda_\nu^{-1}=\sigma_{\nu N}\,\rho$ 
(the conversion from matter density to number density via 
$N_A/{\rm gm}$ is implicit). 
The probability to produce a tau lepton in the interval $dx$ is
$\frac{dx}{\lambda_\nu}$.  The produced $\tau$ carries typically 80\% of the
parent neutrino energy; we approximate this as 100\%.  Then the $\tau$
produced at point $x$ emerges from the surface with energy
$E_\tau=E_\nu\,e^{-(l-x)/\lambda_\tau}$.  The probability of a $\tau$
produced at point $x$ to emerge with sufficient energy $E_{\rm th}$ 
to produce an observable shower can be approximated as 
$P_{\tau \rightarrow {\rm UAS}}=\Theta (\lambda_\tau+x-l)$, with
$\lambda_\tau=\frac{1}{\beta_\tau\,\rho_{sr}}\,\ln(E_\nu/E_{\rm th})$;
$\beta_\tau \approx 0.8 \times 10^{-6} {\rm cm}^2/g$~\cite{beta}
is the exponential energy-attenuation coefficient.  
For taus propagating through rock,
one can take $\lambda_\tau \approx 22$~km for 
$E_\nu \sim 10^{20}$eV and $E_{\rm th}\sim 10^{18}$eV,
while for taus propagating through ocean $\lambda_\tau$ is 2.65 times
larger.  The rates in Fig.~2 are shown for UAS over land.  
An UAS over an ocean, where part or even all of the tau path is 
in water \cite{water}, is best addressed with a simulation. 
The threshold $E_{\rm th}$ depends on details of detector sensitivity and 
aperture, as well as the assumed neutrino spectrum; 
these can suitably be incorporated in a threshold parameter.

Taking the product of these conditional probabilities and integrating over
the interaction site $x$ we get the probability for a tau neutrino
incident along a chord of length $l$ to produce an UCL:
\begin{equation}
\label{probUCL}
P_{{\nu_\tau}\rightarrow\tau}(l)= \int_{l-\lambda_\tau}^l 
\frac{dx}{\lambda_\nu} \, e^{-x/\lambda_\nu}
=(e^{\lambda_\tau/\lambda_\nu}-1) \,e^{-l/\lambda_\nu}\,.
\end{equation}
The emerging tau decays in the atmosphere with probability 
$P_d= 1-\exp(-2 R_\oplus H/c\tau_\tau l) $, where $H\approx 10$~km  
parametrizes the height of the atmosphere.
Thus, the probability for a tau-neutrino to produce an 
UAS is
\begin{equation}
\label{probUAS}
P_{{\nu_\tau}\rightarrow{\rm UAS}}(l)= 
( 1-e^{-2 R_\oplus H/c\tau_\tau l} ) 
\, P_{{\nu_\tau}\rightarrow\tau}(l)\,.
\end{equation}

Next we calculate the probability for an incident neutrino trajectory
to have chord length $l$.
Due to isotropy of the neutrino flux, 
it is enough to consider incident neutrinos with 
parallel trajectories.  
The length $l$ and impact parameter $h$ 
with respect to the Earth's center of a chord are related
by $l^2/4+h^2=R_\oplus^2$.   
The fraction of neutrinos with chord lengths in the
interval $\{l, l+dl\}$ is therefore 
%
\begin{equation}
P_{\rm chord}(l) dl = \frac{2\pi\ h \ dh}{\pi R_\oplus^2}=  
\frac{l}{2 R_\oplus^2} dl .
\label{Pl}
\end{equation}
To get an event rate probability from the incident neutrino flux,
there are two further geometric factors to be included:
the solid angle factor $\pi$ for a planar detector 
with hemispherical sky-coverage, 
and the tangential surface 
area $A$ of the detector \cite{alternative}.

Putting all probabilities together, we arrive at 
the rate of UCL and UAS events:
\begin{equation}
R_{_{\tau({\rm UAS})}} =  F_{\nu_\tau} \pi A  
\int_0^{2R_\oplus} 
\frac{l\,dl}{2\,R_\oplus^2} 
\,P_{{\nu_\tau}\rightarrow{\tau({\rm UAS)}}}(l)\,.
\label{prob2}
\end{equation} 
The double integral in eq.\ (\ref{prob2}) is easily done analytically 
for the UCL case, and for the UAS case when 
the angle above the horizon satisfies 
$\theta \gg (E_\tau/10^{19}{\rm eV})\deg$, such that 
$2\,R_\oplus\,H/c\tau_\tau l \ll 1$.
These results in the limit $\lambda_\nu \gg \lambda_\tau$ are
\begin{equation}
R_\tau =  F_{\nu_\tau} \pi A  \; \frac{1}{2} \xi_\tau\xi_\nu
\left[ 1-e^{-2/\xi_\nu}\left( 1+2/\xi_\nu\right)\right]\,,
\label{prob31}
\end{equation}
where $\xi_\tau = \lambda_\tau/R_\oplus$ and 
$\xi_\nu = \lambda_\nu/R_\oplus$, and
\begin{equation}
R_{_{\rm UAS}} =  F_{\nu_\tau} \pi A  \;
\frac{H\,\xi_\tau}{c\tau_\tau}\,\left(1-e^{-2/\xi_\nu}\right)\,.
\label{prob3}
\end{equation}
For neutrino trajectories through the Earth's mantle,
$\xi_\nu = 0.66/\sigma_{33}$, where $\sigma_{33}$ is 
the neutrino cross section in units of $10^{-33}{\rm cm}^2$.

The tau flux scales as  $\lambda_\nu^{-1}\propto\sigma_{\nu N}$ 
for $\lambda_\nu \gg R_\oplus$ (i.e. small $\sigma_{\nu N}$),
because the large neutrino MFP 
exceeds the Earth's diameter, making the interactions rarer
for increasing $\lambda_\nu$.
For $\lambda_\tau \ll \lambda_\nu \ll R_\oplus$ 
(i.e. $\sigma_{\nu N} \stackrel{>}{_{\scriptstyle \sim}}
  2\times 10^{-33}{\rm cm}^2$), 
the probability scales as 
$\lambda_\nu\propto\sigma_{\nu N}^{-1}$. 
The rise with increasing MFP is attributable to shrinkage of the Earth's
``shadow'' and the consequent increase in the target volume.
The ratio $r_\tau=R_\tau /F_\nu \pi A$ is
shown in Fig.~1.   
We note that the neutrino
cross section may be determinable from the angular distribution of 
UCL events alone (if they can be observed), 
independent of the neutrino flux.
One expects the angular distribution of UCL to peak near 
$\cos \theta_{\rm peak} \sim \lambda_\nu/ 2 R_\oplus$, 
which implies
$\sigma_{\nu N}\sim 
\left( 2\,\langle\rho\rangle\,R_\oplus\,
\cos\theta_{\rm peak}\right)^{-1}\,.$ 

The UAS rate scales differently from the UCL rate,
due to the dependence on path length in the atmosphere.
\begin{figure}[t]
\centering
\hspace*{-5.5mm}
\leavevmode\epsfysize=4cm \epsfxsize=8cm \epsfbox{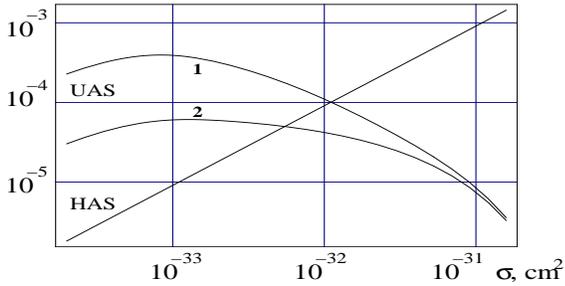} 
\caption[fig. 2]{\label{fig2} The air shower probability per incident tau
neutrino $R_{\rm UAS}/F_{\nu_\tau} \pi A$ as a function of the neutrino
cross section (eq.(6)). 
The incident neutrino energy is $10^{20}$~eV and the assumed 
energy threshold for detection of UAS is
$E_{\rm th}=10^{18}$eV for curve 1 and $ 10^{19}$eV for curve 2.
}
\end{figure}   
In Fig.~\ref{fig2} we show that the the number of expected UAS events per
incoming neutrino as a function of the neutrino cross section
(using the exact integral expression).
The cross section is bounded from below by the value $\sim
2\times 10^{-34}{\rm cm}^2$ measured at HERA at $\sqrt{s}=314$~GeV,
corresponding to a laboratory energy $E_\nu=5.2\times 10^{13}$~eV.  For
comparison, we also show the number of expected HAS events per neutrino
that crosses a 250~km field of view, up to an altitude of 15~km.  
It is clear that for the smaller values of the cross section, UAS events
will outnumber HAS events, while the larger values of the cross section
favor the HAS events.  The ratio of HAS to UAS rates may provide 
a good measure of the cross section. 

We give some examples of the UAS event rates expected from 
a smaller neutrino cross section at $10^{20}$~eV.
Let us choose $\sigma_{\nu N}=10^{-33}{\rm cm}^2$, for example.  
Taking the mantle density of $\rho_m=4.0\,g/{\rm cm}^3$
and $R_\oplus=6.37\times 10^8 {\rm cm}$,
one gets $\xi_\nu=0.65$.
Reference to Fig.\ 1 then shows that the 
$\nu_\tau\rightarrow\tau$ conversion probability is $r_\tau=0.1\%$
for land events with $E_\tau \ge 10^{18}$~eV, 
initiated by a $\sim 10^{20}$~eV primary neutrino.
Including the probability for a tau to decay in the atmosphere,
the $\nu_\tau\rightarrow {\rm UAS}$ probability  
is $4\times 10^{-4}$ ($7\times 10^{-5}$) for a shower-energy threshold 
$E_{\rm th}=10^{18}$~eV ($10^{19}$~eV),
according to Fig.~2.
EUSO and OWL have shower-energy thresholds
$\sim 10^{19}$eV corresponding to curve 2 in Fig.~2. 
They have apertures $\sim 6\times 10^4$km$^2$ and
$3\times 10^5$km$^2$, respectively, for a wide angular-range of UAS.
These detectors should observe $F_{20}$ and $7F_{20}$ UAS events per year, 
respectively (not including duty cycle);
here $F_{20}$ is the incident neutrino flux at 
and above $10^{20}$~eV in units of km$^{-2}$sr$^{-1}$yr$^{-1}$,
one-third of which are $\nu_\tau$'s. 
Including showers from taus originating outside the 
field of view, and direct tau events, increases these rates.
The rates may be further increased in space-based detectors by
tilting towards the horizon so as to maximize the
acceptance for events with smaller chord lengths 
(where the neutrino attenuation is less and the field of view 
is greater, but the energy threshold is higher) 
and to allow more atmospheric path length for tau decay.   
The rates will also increase if $E_{\rm th}$ can be reduced;
reducing $E_{\rm th}$ to $10^{17}$~eV or less 
would allow the GZK flux to be observed and measured.  
Although the spectrum of UHE neutrinos is not known, a variety of assumed 
spectra can be encompassed by our parametrization, and our 
conclusions are not sensitive to spectral details.

Finally, we comment on two important inferences.
First, the reported bounds on the UHE neutrino flux due to the
non-observation of neutrino-initiated HAS~\cite{FlyEye} and 
of radio signals produced by neutrino interactions near the surface of the
moon~\cite{GLN99} are weaker if the cross section is smaller. 
Concerning the lunar radio bound, the lunar radius
is 1740 km, about 3.5 times smaller than that of Earth, and the density of
the Moon is about the same as the Earth's surface density.  Thus,
$\xi_\nu=\lambda_\nu/R$ for the moon is $3.6/\sigma_{33}$, which is about
5.5 times the value for earthly neutrinos.  Consequently, the range of
$\xi_\nu$ for the Moon allowed by our ignorance of the true neutrino cross
section is very large, and the true neutrino flux limit from lunar radio
could be different from that previously reported.  The second inference has
to do with the predicted angle-independence for upgoing $\tau$-neutrinos at
$\sim 10^{14}$~eV~\cite{HS98}.  For a smaller cross section, a harder
spectrum of unattenuated $\nu_\tau$'s above $10^{14}$~eV and a larger angle
dependence may be expected.

To conclude, the overall prospects for UHE neutrino astronomy are not
diminished by the theoretical uncertainties in value of the
neutrino-nucleon cross section.  UHECR neutrinos can be observed at a
healthy rate for any allowed value of the cross section.  Furthermore,
future neutrino cosmic-ray experiments can determine the neutrino-nucleon
cross section at energies as high as $10^{11}$~GeV, or higher, by comparing
the rates of UAS with those of HAS; or by measuring the angular
distribution of UCL events.

We thank K.~Arisaka, Z.~Bern, D.~Bodeker, J.~Feng, F.~Halzen, G.~Sigl,
G.~Sterman, and F.~Wilczek for helpful discussions.  This work was
supported by the DOE grants DE-FG03-91ER40662 (AK) and DE-FG05-85ER40226
(TJW).



\begin{references}
\bibitem{water}

The entire path length in earth of a tau emerging at an angle 
$\lsim 4^\circ$ relative to the horizon is typically in water. 
The neutrino path length at $4^\circ$  is $\sim 900$~km, 
equal to the MFP in mantle of a neutrino with  
$\sigma_{\nu N} = 5\times 10^{-33}{\rm cm}^2$.
The shower probabilities 
of such low-angle ocean events will be
lower for $\sigma_{33} < 5 $ and a factor of a few higher for 
$\sigma_{33} > 5 $ 
than the probabilities shown in Fig.~2.
\bibitem{alternative}
Of course, the same probability measure 
can also be  
obtained by taking the product 
of the solid-angle element $d\Omega=2\,\pi\,d\cos\theta_z$
times the projected area of the detector $A\,\cos\theta_z$.
The result is $2\,\pi\,A\,\cos\theta_z\,d\cos\theta_z$,
which equals $(\pi A l)/(2 R^2_\oplus) dl$ when 
use is made of $l=2\,R_\oplus\,\cos\theta_z$.
\end{references}
\end{document}